\documentclass[a4paper,11pt]{article}
\usepackage[utf8]{inputenc} 
\usepackage[T1]{fontenc} 
\usepackage{geometry}
\usepackage{setspace}
\usepackage{authblk}

\usepackage[hidelinks]{hyperref} 
\usepackage{url} 
\usepackage{graphicx}
\usepackage[shortlabels]{enumitem}
\usepackage{caption}
\usepackage[scaled=1]{helvet}

\usepackage{pdflscape}
\usepackage{longtable} 
\usepackage{threeparttablex}
\usepackage{booktabs} 
\usepackage{lscape} 
\usepackage{array} 
\usepackage{multirow}
\usepackage{colortbl}
\usepackage{xcolor}
\usepackage{amsfonts} 
\usepackage{dsfont} 
\usepackage{microtype} 
\usepackage{lipsum} 
\usepackage{doi}
\usepackage{framed}
\usepackage{longtable}
\usepackage{ragged2e}
\usepackage{amsmath,amssymb}
\usepackage{calc}
\usepackage{mathtools}
\usepackage{fancyhdr}
\usepackage{siunitx}

\usepackage{tikz}
\usetikzlibrary{arrows.meta, positioning, calc}

\usepackage[
  backend=biber,
  style=numeric-comp,      
  sorting=none,
  maxbibnames=6,
  minbibnames=3,           
  giveninits=true,
  terseinits=true,         
  autocite=superscript,    
  doi=false,
  isbn=false,
  url=false,
  eprint=false
]{biblatex}

\AtEveryBibitem{%
  \ifentrytype{techreport}{%
    \printfield{url}%
    \printurldate
  }{}%
}

\usepackage{xr-hyper}
\graphicspath{}
\author[1,*]{Marc Delord}
\affil[1]{Department of Primary Care \& Data Science,\\
School of Life Course \& Population Health Sciences, King's College London, London, UK}
\affil[*]{Correspondence: marc.delord@kcl.ac.uk\\ Tel: +44 20 7848 8710}

\title{\Large{Non-Invariance in Nested Prediction Models\\ under Selective Predictor Availability}}
\date{}

\begin{document}

\maketitle

\setstretch{1.3}
\begin{abstract}
\noindent
We use nested models as a framework for characterising the consequences of a selectively measured predictor in clinical prediction models. In this framework, a model containing predictors available in the target population is referred to as the restricted model, while the extended model additionally includes a selectively measured predictor. We show that non-invariance in the restricted model between selected patients and the target population decomposes into components due to omission of the additional predictor, potential residual non-invariance, and their interaction, and that it requires selection through manifestations of the disease process beyond the restricted-model predictors. We further show that imputation based on the conditional distribution of the additional predictor in the selected population transfers the restricted-model non-invariance to the imputed extended model as imputation bias. We propose diagnostic tests for selective measurement and restricted-model non-invariance that can be applied to routinely collected data, and illustrate the framework using publicly available intensive care data, in which blood lactate is selectively measured. A simulation study with binary and time-to-event outcomes illustrates these results and evaluates the diagnostic tests in finite samples. The framework provides a formal basis for understanding how selective predictor measurement affects generalisability of clinical prediction models using routinely collected health data.
\end{abstract}

\subsubsection*{Keywords:}
Clinical prediction models; Non-invariance; Model misspecification; Selection bias; Imputation; Electronic health records

\setstretch{1.3}
\newcounter{as}
\newcommand{\astag}{\refstepcounter{as}\tag{A\theas}}

\newcounter{res}
\newcommand{\restag}{\refstepcounter{res}\tag{R\theres}}

\newcounter{dag}
\newcommand{\dagtag}{\refstepcounter{dag}\tag{D\thedag}}

\newcounter{eq}
\newcommand{\eqtag}{\refstepcounter{eq}\tag{Eq\theeq}}

\newcounter{mod}
\newcommand{\modtag}{\refstepcounter{mod}\tag{M\themod}}

\section*{Introduction}

Selective measurement occurs whenever the decision to record a clinical predictor depends on a patient's own characteristics or disease process, rather than on chance alone \autocite{Goldstein2016,goldstein2017opportunities}. This is the rule rather than the exception in routinely collected health data: laboratory tests are ordered in response to clinical suspicion, symptoms, or prior results, so predictor availability itself carries information about the underlying disease process. When such a predictor is used in a clinical prediction model and its measurement depends on manifestations of the disease process that are not included in the model, predictor availability becomes structurally associated with outcome risk conditional on the model's predictors, and the model becomes prone to a form of selection bias in the subset of patients for whom the predictor is available \autocite{Hernn2004,lu2022toward,lu2024selection}.\\

Prediction models are widely used in the clinical setting to estimate an individual's risk of future health outcomes from routinely collected characteristics \autocite{altman2000we, Moons2009a, steyerberg2009applications, steyerberg2013prognosis}, and their use implicitly assumes that predictors are available at the point of care in the target population \autocite{steyerberg2013prognosis,moons2019probast,efthimiou2024developing}.In practice, however, as noted above, predictor availability is rarely random, and model development, validation, and application are consequently restricted to selected subsets of patients rather than the full target population \autocite{sisk2021informative,wells2013strategies,madden2016missing}.\\

As different sets of predictors may select different, heterogeneous subsets of a target population, this raises the central question addressed in this paper: how does selection by predictor availability affect the generalisability of a clinical prediction model to its intended target population \autocite{pearl2011transportability,collins2024evaluation}? While informative measurement processes have been recognised in routinely collected data \autocite{sisk2021informative,wells2013strategies,madden2016missing,Goldstein2016}, the full implications for the generalisability of clinical prediction models developed or validated in selected subsets of patients have yet to be formally characterised.\\

A clear instance of this situation is provided by the Kidney Failure Risk Equation (KFRE) \autocite{tangri2011predictive}. This model aims at predicting progression to kidney failure in patients with chronic kidney disease. The three-variable KFRE is based on age, sex, and estimated glomerular filtration rate (eGFR), whereas the four-variable model additionally includes the urine albumin-to-creatinine ratio (ACR). In routine care, eGFR is commonly measured, while ACR is less consistently recorded and is often assessed in response to clinical indication \autocite{fraser2015timeliness,chu2023estimated,ketema2025quality}. As a result, the four-variable model is effectively restricted to a subgroup of patients with available ACR records, with a risk profile that may differ from that of patient with chronic kidney disease (CKD) considered as the KFRE target population.

The same situation arises in intensive care, where routine variables such as age, vital signs and basic laboratory tests are recorded for almost all patients, whereas blood lactate, a strong marker of hypoperfusion and mortality, is measured when clinicians suspect circulatory failure or sepsis. We use this setting, with publicly available intensive care data \autocite{silva2012predicting}, to illustrate the proposed framework.\\

We formalise this setting using nested prediction models under differential predictor availability where $X$ represents a set of predictors available in the target population and $Z$ an additional predictor selectively measured. A restricted model based on $X$ can be estimated in the target population, whereas the extended model based on $(X,Z)$ is estimated in the selected subset of patients based on availability of $Z$ in the target population. We used the directed acyclic graph (DAG) framework to illustrate the causal structures underlying the proposed nested prediction models under differential predictor availability and interpret the restricted-model non-invariance as a consequence of model misspecification arising from omission of additional predictors \autocite{heckman1979sample}.\\

This paper is organised as follows. In the first section, we introduce the formal framework of two nested prediction models under differential predictor availability. In the second section, we derive a decomposition of restricted-model non-invariance and distinguish internal and external routes of selection. In the third section, we show how, under basic assumptions, imputation of the additional predictor in the non-selected population translates restricted-model non-invariance into unobserved imputation bias. The fourth section proposes diagnostic tests that can be applied to routinely collected data. The fifth section evaluates the framework empirically, in a simulation study with binary and time-to-event outcomes and in an application to publicly available intensive care data, and the final section discusses the implications of our results.

\section*{Nested prediction models under differential predictor availability}
\subsection*{Clinical setting}

We start by formalising the setting of nested prediction models mentioned above, defined in a specified target population. Let \(Y\) denote the outcome, \(X = (X_1, X_2, \dots)\) a set of routinely available predictors defining the restricted model, and \(Z\) an additional predictor required to form the extended model. Let $S$ denote the indicator of availability of $Z$, with $S=1$ when $Z$ is observed and $S=0$ otherwise.  We also consider $U$, the underlying unobserved disease process, $W$, an observable proxy for \(U\), such as disease symptoms or recorded risk factors not included in the nested prediction models. The decision to measure $Z$ (i.e. setting $S=1$) can be triggered by both values of the measured predictors ($X$) or other observable manifestations of the underlying disease process $W$.\\

This clinical setting can be represented by a directed acyclic graph (DAG), as shown in DAG~(\ref{dag_0}):

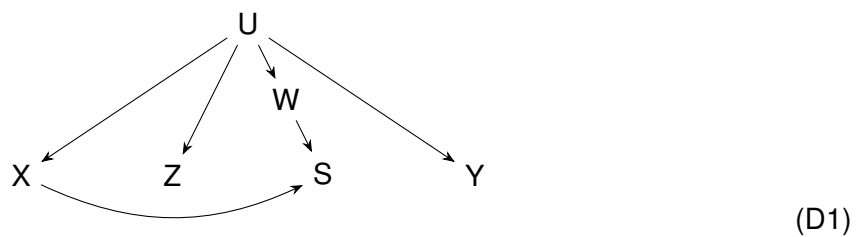
\begin{equation}
\dagtag
\label{dag_0}
\begin{tikzpicture}[->, >=Stealth]

\node (U) at (0,-1) {U};
\node (W) at (.5,-2) {W};

\node (X) at (-3,-3) {X};
\node (Z) at (-1,-3) {Z};
\node (S) at (1,-3) {$\text{S}_\text{}$};
\node (Y) at (3,-3) {Y};

\draw (U) -- (X);
\draw (U) -- (Z);
\draw (U) -- (Y);

\draw (X) to[bend right=25] (S);
\draw (U) -- (W);
\draw (W) -- (S);

\end{tikzpicture}
\end{equation}
In this representation, the outcome $Y$ depends on the underlying disease process $U$ only, so that $X$ and $Z$ are prognostic without being causes of $Y$.

\subsection*{Selectively measured predictors}

Selective measurement occurs when the decision to measure a clinical predictor depends on patient information ($\mathcal I$), rather than occurring completely at random \autocite{Goldstein2016}. In the clinical setting of DAG \ref{dag_0}, this corresponds to  $S$ being dependent on predictors $X$ or other manifestations of the underlying disease process $W$, noted $S \not\!\perp\!\!\!\perp \mathcal I$ with \(\mathcal I=(X,W)\), or equivalently

\[
\mathrm{ P }( S = 1 \mid \mathcal I ) \neq \mathrm{ P }( S = 1 ).
\]

Of note, $Z$ being measured completely at random, i.e. \[S \perp\!\!\!\perp (Z,\mathcal{I})\] implies

\begin{equation}
\astag
\label{a0}
Z \perp\!\!\!\perp S \mid  X.
\end{equation}

Conversely, absence of conditional independence of $Z$ and $S$ given $X$ is noted

\begin{equation}
\tag{$\neg$ \ref{a0}} Z \not\!\perp\!\!\!\perp S \mid X,
\end{equation}

or equivalently,

 \[
 \tag{$\neg$ \ref{a0}}
 P(Z\mid  X,S=1)\neq P(Z\mid X).
 \]

Conditions \ref{a0} and $\neg$ \ref{a0} correspond respectively to missing-at-random (MAR) and missing-not-at-random (MNAR) mechanisms for $Z$, conditional on $X$, in the terminology of the missing-data literature \autocite{vanBuuren2018}.\\

In DAG \ref{dag_0}, $\neg$~\ref{a0} holds, although $Z \perp\!\!\!\perp S \mid \mathcal I$, since $Z$ and $S$ are associated through $U$ and $W$; when measurement depends on $X$ only, \ref{a0} holds.\\

We also define predictor non-invariance between the selected and target populations through the conditional distribution of $Z$:

\[
\delta_{Z\mid  X}
=
P(Z\mid  X,S=1)-P(Z\mid  X).
\]

In addition, $Z$ is said to carry residual prognostic information beyond the restricted-model predictors if

\begin{equation}
\astag
\label{a1}
Z \not\!\perp\!\!\!\perp Y \mid  X.
\end{equation}

In the proposed setting, selective measurement of $Z$, leading to conditional dependence between $Z$ and $S$ given $X$ ($\neg$ \ref{a0}), becomes relevant for prediction through the structural condition of conditional dependence between $Z$ and the outcome given $X$ (\ref{a1}).

\subsection*{Distributional generalisability}

Let the selected population be defined by \(\{S=1\}\) and the target population by \(\{S\in\{0,1\}\}\), such that

\[
\{S=1\}\subset\{S\in\{0,1\}\}.
\]

We define distributional generalisability as invariance of the conditional outcome distribution between the selected and target populations, that is,

\begin{equation}
\astag
\label{a2}
\mathrm{P}(Y \mid X, Z, S = 1)
=
\mathrm{P}(Y \mid X, Z).
\end{equation}

In this setting, the core inferential question becomes: is a model estimated in selected patients generalisable to the target population? In addition to model invariance across populations \autocite{pearl2011transportability}, this question can also be considered in terms of selection bias arising from specification error due to a selectively measured predictor \autocite{heckman1979sample}.\\

Assuming that \(Z\) is selectively measured,  non-invariance in the extended model

\[
\delta_{X,Z}
=
\mathrm{P}(Y \mid X, Z, S = 1)
-
\mathrm{P}(Y \mid X, Z)
\]

is also the conditional selection bias. Practically, this corresponds to a model being estimated in patients with available values of \(Z\), i.e. \(\{S=1\}\), while \(\mathrm{P}(Y \mid X, Z, S = 0)\), and consequently \(\mathrm{P}(Y \mid X, Z)\), remain unobserved, with

\begin{equation}
\eqtag
\label{eq1}
\mathrm{P}(Y \mid X, Z)
=
\pi_{X,Z}\,\mathrm{P}(Y \mid X, Z, S = 1)
+
(1-\pi_{X,Z})\,\mathrm{P}(Y \mid X, Z, S = 0),
\end{equation}

where

\[
\pi_{X,Z}=\mathrm{P}(S=1\mid X,Z).
\]

Finally, we obtain the marginal selection bias by averaging the conditional selection bias over the conditional distribution of $Z$ in the target population:

\[
\Delta_{X,Z} =
\int
\delta_{X,Z}\,
d\mathrm{P}(Z \mid X).
\]

\subsection*{Selection-related prognostic information and non-invariance in the restricted model}

\subsubsection*{Invariance in the extended model}

To better understand and interpret the impact of selective measurement of a predictor in a prediction model, we consider the situation in which the extended model is invariant whereas the restricted model is non-invariant, i.e.

\[
\mathrm{P}(Y\mid X,Z,S=1)=\mathrm{P}(Y\mid X,Z) \quad \text{and} \quad \mathrm{P}(Y\mid X,S=1) \neq \mathrm{P}(Y\mid X).
\]

Integrating over \(Z\) in the restricted model across the selected and the target population yields

\[
\mathrm{P}(Y\mid X,S=1)
=
\int
\mathrm{P}(Y\mid X,Z,S=1)\,
d\mathrm{P}(Z\mid X,S=1),
\]

and

\[
\mathrm{P}(Y\mid X)
=
\int
\mathrm{P}(Y\mid X,Z)\,
d\mathrm{P}(Z\mid X).
\]

Since the extended model is invariant,

\begin{align*}
\mathrm{P}(Y\mid X,S=1)
&=
\int
\mathrm{P}(Y\mid X,Z,S=1)\,
d\mathrm{P}(Z\mid X,S=1)\\
&=
\int
\mathrm{P}(Y\mid X,Z)\,
d\mathrm{P}(Z\mid X,S=1),
\end{align*}

which under non-invariance in the restricted model implies

\begin{equation}
\mathrm{P}(Z\mid X,S=1)
\neq
\mathrm{P}(Z\mid X)
\tag{$\neg$ \ref{a0}}
\end{equation}

or equivalently,

\begin{equation}
Z \not\!\perp\!\!\!\perp S \mid X.
\tag{$\neg$ \ref{a0}}
\end{equation}

Finally, the restricted model being non-invariant while the extended model is invariant also implies that

\begin{equation}
Z \not\!\perp\!\!\!\perp Y \mid X,
\tag{\ref{a1}}
\end{equation}

that is, $Z$ carries residual prognostic information.\\

To summarise, under extended-model invariance, restricted-model non-invariance implies $\neg$~\ref{a0} and \ref{a1} (see Appendix \ref{ap2} for a formal proof):

\begin{equation*}
\left.
\begin{array}{c}
\underbrace{\mathrm{P}(Y\mid X,Z,S=1)=\mathrm{P}(Y\mid X,Z)}_{\text{extended-model invariance (\ref{a2})}}\\[1.5ex]
\underbrace{\mathrm{P}(Y\mid X,S=1) \neq \mathrm{P}(Y\mid X)}_{\text{restricted-model non-invariance}}
\end{array}
\right\}
\Rightarrow
\underbrace{Z \not\!\perp\!\!\!\perp S \mid X}_{\text{$\neg$\,\ref{a0}}}
\quad \text{and} \quad
\underbrace{Z \not\!\perp\!\!\!\perp Y \mid X}_{\text{\ref{a1}}}
\end{equation*}

~\\
In this situation, non-invariance in the restricted model reflects model misspecification resulting from omission of a predictor
\(Z\) carrying residual prognostic information, while the inclusion of \(Z\) restores invariance in the extended model. This corresponds to the two structural conditions identified above.\\

Alternatively, non-invariance observed in the restricted model and resolved by inclusion of $Z$ can be interpreted as $Z$ fully capturing the selection-related prognostic information beyond $X$. In other words, although the decision to measure $Z$ selects patients whose conditional outcome distribution differs from that of the target population given $X$, conditioning on $(X,Z)$ accounts for this difference.

However, if the extended model remains non-invariant after inclusion of \(Z\), this indicates that \(Z\) does not fully capture the selection-related prognostic information beyond \(X\). Equivalently, the selection mechanism remains prognostically informative after conditioning on the predictors included in the extended model, that is,

\[
 S \not\!\perp\!\!\!\perp Y \mid X , Z,
\]

which is equivalent to

\begin{equation*}
\mathrm{P}(Y \mid X,  Z ,S=1)
\neq
\mathrm{P}( Y \mid X , Z).
\end{equation*}

This further suggests that non-invariance in the restricted model can be decomposed into a component explained by $Z$ and a residual component.

\subsubsection*{Non-invariance in the extended model}

We now consider non-invariance in the restricted model while relaxing the hypothesis of invariance in the extended model as discussed above.

Non-invariance in the restricted model conditional on availability of \(Z\) can be expressed as

\[
\delta_{X,S} =
\mathrm{P}(Y \mid X,S=1)-\mathrm{P}(Y \mid X).
\]

After integrating over the additional predictor \(Z\) (see Appendix \ref{ap3}), we obtain

\[
\Delta_{X,S}
=
\Delta_{Z\mid X}^{\mathrm{exp}}
+
\Delta_{Z , X}^{\mathrm{res}}
+
\Delta_{Z , X}^{\mathrm{int}},
\]
where

\[
\begin{cases}
\Delta^{\rm exp}_{Z \mid X}
=
\int
\mathrm{P}(Y \mid X,Z)\,
d \delta_{Z \mid X}
\\
\Delta^{\rm res}_{Z , X}
=
\int
\delta_{X,Z}\,
d\mathrm{P}(Z \mid X)
\\
\Delta^{\rm int}_{Z , X}
=
\int
\delta_{X,Z}
\,
d\delta_{Z \mid X}.
\end{cases}
\]

Under $\neg$~\ref{a0} (predictor non-invariance) and \ref{a1} (predictor-associated residual prognostic information), the first term represents the non-invariance attributable to omission of \(Z\), and resolved by \(Z\) in the extended model; the second term represents the marginal residual non-invariance; and the third term represents the interaction between predictor non-invariance and residual non-invariance. Of note, the second term represents also the marginal selection bias in the extended model.\\

The interaction term vanishes when $\delta_{X,Z}$ does not depend on $Z$, that is, when the conditional selection bias is homogeneous across the values of the selectively measured predictor $Z$. In practice, this interaction is expected to be negligible when this condition is approximately satisfied. Neglecting this interaction yields the first-order approximation

\[
\Delta_{X,S}
\approx
\Delta_{Z\mid X}^{\mathrm{exp}}
+
\Delta_{Z, X}^{\mathrm{res}}.
\]

For the remainder of this paper, we will assume homogeneous conditional selection bias, i.e.

\[
\delta_{X,Z} = c(X) \quad \text{for all $Z$}.
\]

\subsection*{Invariance in the restricted model}

From the decomposition of non-invariance in the restricted model, restricted-model invariance, $\Delta_{X,S}=0$, arises in two cases: either both components are null, $\Delta_{Z \mid X}^{\mathrm{exp}} = 0$ and $\Delta_{Z , X}^{\mathrm{res}} = 0$, or they cancel, $\Delta_{Z , X}^{\mathrm{res}} = -\Delta_{Z \mid X}^{\mathrm{exp}} \neq 0$.\\

The first case arises under two conditions. Under \ref{a0} (predictor invariance given $X$), $\delta_{Z\mid X}=0$, so that $\Delta_{Z \mid X}^{\mathrm{exp}} = 0$ and, since $\Delta_{X,S}=0$, $\Delta_{Z , X}^{\mathrm{res}} = 0$: adding $Z$ does not induce non-invariance in the extended model. Under $\neg$~\ref{a0} and $\neg$~\ref{a1} (absence of residual prognostic information), the predictor non-invariance is not associated with the outcome, so that again both components are null.\\

The second case arises under $\neg$~\ref{a0} and \ref{a1}: the additional predictor is non-invariant given $X$ and carries residual prognostic information, so that $\Delta_{Z \mid X}^{\mathrm{exp}} \neq 0$ in general, and restricted-model invariance requires $\Delta_{Z , X}^{\mathrm{res}} = -\Delta_{Z \mid X}^{\mathrm{exp}}$; in this case, the inclusion of $Z$ induces selection bias in the extended model.\\

Of note, exact cancellation of the effects of selection on the conditional distribution of $Z$, when integrated over the outcome risk, may result in a null explained non-invariance ($\Delta^{\rm exp}_{Z\mid X}=0$) despite an underlying selection mechanism ($\neg$~\ref{a0} and~\ref{a1}); such cancellation is possible but non-generic.

\subsection*{Internal and external routes of selection}

We distinguish two routes of selection in DAG~(\ref{dag_0}). First, selection operates through an internal route when the decision to measure $Z$ depends on the restricted-model predictors $X$. This route is represented by $X$ being a parent of $S$ ($U \rightarrow X \rightarrow S$). The second route of selection is external given $X$, i.e. when the decision to measure $Z$ depends on other manifestations of the underlying disease process ($U \rightarrow W \rightarrow S$).\\

Under extended-model invariance (\ref{a2}), restricted-model non-invariance requires $\neg$~\ref{a0} and \ref{a1}. Under the internal route, the paths between $S$ and $Z$, and between $S$ and $Y$, go through $X$ and are blocked given $X$ \autocite{greenland1999causal,pearl2016causal,hernan2020causal}, so that \ref{a0} holds and both the extended and the restricted models are invariant, $\Delta_{X,S}=0$. This corresponds to the known result that selection depending only on the predictors of a model does not bias the conditional distribution of the outcome given these predictors \autocite{little1992regression,white2010bias,bartlett2015asymptotically,bareinboim2014recovering}. This holds whatever the strength of selection on $X$, and whether or not $Z$ carries residual prognostic information (\ref{a1}). Selection on the restricted-model predictors thus changes which patients are selected, but not their risk given $X$.\\

Under the external route, the path $S \leftarrow W \leftarrow U \rightarrow Z$ is open given $X$, so that $\neg$~\ref{a0} holds, and when $Z$ carries residual prognostic information (\ref{a1}), the restricted model, from which $Z$ is omitted, is non-invariant, this non-invariance being entirely explained by the omission of $Z$.\\

When $Z$ is only a partial measure of $U$ given $X$, the external route also leaves the path $S \leftarrow W \leftarrow U \rightarrow Y$ open given $X$ and $Z$, so that the extended model is itself non-invariant and residual non-invariance adds to the explained component; conversely, if $X$ and $Z$ together fully captured $U$, conditioning on $(X,Z)$ would block this path and the extended model would be invariant.\\

Accordingly, restricted-model non-invariance requires an external route of selection, while selection on the restricted-model predictors alone, however strong, does not induce non-invariance. Of note, a difference in the distribution of $X$ between selected and non-selected patients therefore indicates selection, but not non-invariance.\\

Finally, when several manifestations of the disease process drive measurement, such as $W_1$ and $W_2$, availability of $Z$ is a collider between them, which become associated in selected patients. The conclusion above is unchanged, since each external route induces non-invariance, and $W$ is then tested jointly in the diagnostic test described below.

\section*{Implications for prediction model validation}

Imputation methods consist of using the conditional distribution of the predictor estimated in selected patients to predict its missing values in non-selected patients \autocite{vanBuuren2018}. In our setting, and considering only predictors included in the extended model, this consists of using the conditional distribution of $Z$ estimated in selected patients to predict its missing values:

\[
\tilde{Z} =
\begin{cases}
Z, & S=1,\\
\hat{Z}, & S=0,
\end{cases}
\]

where

\[
\hat{Z} \mid X,S=0 \sim  \textrm{P}(Z \mid X,S=1),
\]

or equivalently
\begin{equation}
 \textrm{P}(\hat Z\mid X,S=0)
=
 \textrm{P}(Z\mid X,S=1).
\label{a4}
\astag
\end{equation}

After the imputation step, predictions are computed using the predictive distribution of the outcome estimated in the selected population using imputed values of $Z$:

\begin{equation}
\mathrm{P}(Y \mid X,\hat{Z},S=0) = \mathrm{P}(Y \mid X,Z,S=1).
\label{a5}
\astag
\end{equation}

However, when selection affects the conditional distribution of $Z$ ($\neg$ \ref{a0}), the imputation process incorrectly assumes the conditional distribution of $Z$ in selected patients \( \textrm{P}(Z\mid X,S=1)\), rather than the conditional distribution \( \textrm{P}(Z\mid X,S=0)\), for patients with missing predictor.\\

The consequence of using the conditional distribution of $Z$ in selected patients to impute $Z$ in non-selected patients when $Z$ is selectively measured can be explored by expressing the difference (or bias) between the predictive distribution of the outcome after imputation and the true unknown predictive distribution of the outcome in the non-selected population

\[
\delta^{\rm imp}_{ Z \mid X, S=0}
=
\hat{\textrm{P}}(Y\mid X , S=0)
-
 \textrm{P}(Y\mid X,S=0).
\]

Since imputation is performed only in non-selected patients, the imputation bias over the target population is $\Delta^{\rm imp}_{Z\mid X}=(1-\pi_X)\,\delta^{\rm imp}_{Z\mid X,S=0}$, with $\pi_X=\mathrm P(S=1\mid X)$. After integrating the first term over $\hat{Z}$ and the second over $Z$ (see Appendix \ref{ap4}), we obtain:

\[
\Delta^{\rm imp}_{Z\mid X}
=
\Delta^{\rm exp}_{Z\mid X}
+
\Delta^{\rm res}_{Z , X}
=
\Delta_{X,S}.
\]

This final result shows that non-invariance in the restricted model is fully transferred to the imputed extended model in the form of imputation bias. Of note, the equality $\Delta^{\rm imp}_{Z\mid X}=\Delta_{X,S}$ does not require the homogeneity assumption, which only removes the interaction term from its decomposition (Appendix \ref{ap4}). The Appendix also derives the identity between conditional imputation bias and conditional non-invariance in the extended model (Appendix \ref{ap6}).

\section*{Diagnostic tests}

The proposed framework suggests empirical diagnostic tests that can be applied to routinely collected data, when $Z$ is selectively observed ($S=1$).\\

\textit{Selective measurement through the restricted-model predictors.} Whether availability of $Z$ depends on $X$ can be assessed under the logistic regression model

\begin{equation}
\label{mo_2}
\modtag
\mathrm{logit}\{\mathrm{P}(S=1\mid X)\} = \gamma_0+\gamma_X^\top X,
\end{equation}
by testing $H_0:\gamma_X=0$. Since selection on $X$ alone leaves both models invariant, this test characterises selection but does not, by itself, indicate non-invariance.\\

\textit{Selective measurement through other manifestations of the disease process.} Whether availability of $Z$ depends on $W$ conditional on $X$, corresponding to an external route of selection, can be assessed under
\begin{equation}
\label{mo_3}
\modtag
\mathrm{logit}\{\mathrm{P}(S=1\mid X,W)\} = \gamma_0+\gamma_X^\top X+\gamma_W^\top W,
\end{equation}
by testing $H_0:\gamma_W=0$. These two tests do not assess condition A1 directly, since $Z$ is not observed in non-selected patients, but evidence that selection depends on $W$ given $X$ makes A1 implausible.\\

\textit{Residual prognostic information.} Condition A2 can be assessed in selected patients by comparing the restricted and extended models, for instance with a likelihood ratio test of $H_0:\beta_Z=0$ in
\begin{equation}
\label{mo_A2}
\modtag
g\{E(Y\mid X,Z,S=1)\} = \beta_0+\beta_X^\top X+\beta_Z Z.
\end{equation}
Under~\ref{a2}, $\mathrm P(Y\mid X,Z)$ is the same in selected patients and in the target population, so that residual prognostic information detected in selected patients carries over to the target population.\\

\textit{Restricted-model non-invariance.} Non-invariance can be estimated by applying the restricted model estimated in selected patients to non-selected patients and comparing the mean predicted and observed risks,
\[
\hat\Delta_{S=0} = \frac{1}{n_0}\sum_{i:S_i=0}\hat{\mathrm P}(Y_i=1\mid X_i,S=1) - \frac{1}{n_0}\sum_{i:S_i=0} Y_i ,
\]
with uncertainty quantified by bootstrap, resampling selected and non-selected patients separately. This quantity estimates non-invariance at the level of non-selected patients, which is related to $\Delta_{X,S}$ through the factor $(1-\pi_X)$ (Appendix~\ref{ap4}), and equals the imputation bias in non-selected patients derived above.\\

Due to the non-collapsibility of logistic and Cox models, restricted-model non-invariance is assessed on the probability scale rather than through the coefficients of models including the selection indicator $S$ and its interaction with the restricted-model linear predictor \autocite{greenland1999confounding,Hernan2010}. For time-to-event outcomes, the observed risk comes from Kaplan--Meier (or Aalen--Johansen under competing risks) at horizon $t$. However, since proportional-hazards models are generally misspecified when prognostic information is omitted \autocite{gail1984biased,struthers1986misspecified}, the predicted risk is preferably obtained from a direct model with inverse probability of censoring weights \autocite{scheike2008predicting} (Supplementary Material).

\section*{Empirical evaluation}

The framework is evaluated in two complementary ways. A simulation study, in which the data-generating mechanism and the true quantities are known, assesses the identity and the diagnostic tests in finite samples, for binary and time-to-event outcomes. An application to publicly available intensive care data illustrates the diagnostic tests and the estimation of non-invariance and imputation bias in routinely collected data.

\subsection*{Simulation study}

\subsubsection*{Simulation design}

The finite-sample behaviour of the identity and of the diagnostic tests was assessed in a simulation study, designed and reported following Morris et al.\ \autocite{morris2019using}; details are given in the Supplementary Material. Data were generated according to DAG~(\ref{dag_0}). A latent disease process $U \sim N(0,1)$ determined the outcome together with an independent characteristic $X_2$, such as age, and was observed only through its manifestations: a restricted-model predictor $X_1 = U + \varepsilon_1$, another manifestation $W = U + \varepsilon_2$ not included in the models, and the selectively measured predictor $Z = \lambda_Z U + \varepsilon_3$, with $\lambda_Z = 0$ (uninformative) or $0.8$. Measurement of $Z$ followed $\mathrm{logit}\,\mathrm P(S=1) = -0.2 + \gamma_X X_1 + \gamma_W W$, defining five scenarios: no selection, internal route only, external route only, and both routes with moderate or strong external selection. Binary outcomes were generated from a logistic model and time-to-event outcomes from a Weibull model with independent censoring, the quantity of interest being the risk by five years. Each of the 1,000 replicates per scenario contained 2,000 patients, about half of whom were selected.

In each replicate, the restricted and extended models were fitted in selected patients, $Z$ was imputed in non-selected patients from a linear model in $X$, and restricted-model non-invariance and imputation bias were estimated as in the illustration below, with distributional and conditional-mean imputation. Distributional imputation averages each patient's predicted risk over the conditional distribution of $Z$ given $X$, as required by the identity; conditional-mean imputation, which replaces $Z$ by its fitted mean, is included as common practice. Because $Z$ is known for every simulated patient, the extended model could also be evaluated with the true values of $Z$ in non-selected patients; this oracle quantity is the residual component of the decomposition. For the time-to-event outcome, both models targeted the five-year risk directly, by binary regression with inverse probability of censoring weights, and proportional-hazards Cox models were fitted for comparison.

\subsubsection*{Simulation results}

Distributional imputation bias reproduced restricted-model non-invariance in all scenarios, for both outcomes (Table~\ref{tab:sim}): the difference ranged from $-0.01$ to $0.03$ percentage points for the binary outcome and from $-0.03$ to $0.03$ points for the time-to-event outcome, within Monte Carlo error. This held although the extended model was itself not invariant whenever the external route operated, consistent with the derivation of the identity, which relies on the imputation assumptions only. Selection through the restricted-model predictor alone produced no non-invariance, despite strong evidence of selection on $X$. The external route produced non-invariance of about five percentage points for the binary outcome. When $Z$ was uninformative, this non-invariance was entirely residual, the extended model with the true values of $Z$ being equally non-invariant; when $Z$ reflected the disease process, about half of it was explained by $Z$ (for instance, a non-invariance of 5.25 points and a residual non-invariance of 2.79 points with the external route only), and imputation transferred both components. Conditional-mean imputation departed from distributional imputation by 0.7 to 0.8 points, including in the absence of selection. The three diagnostic tests kept their nominal level when the tested effect was absent and detected it in all replicates when it was present; M1 also rejected when selection operated through the internal route only, which produced no non-invariance, whereas M2 identified the external route. For the time-to-event outcome, Cox models gave similar results, their misspecification affecting the restricted and extended models alike; the direct models do not rely on such cancellation.

\IfFileExists{sim_table_main.tex}{\begin{landscape}
\begin{table}[!ht]
\centering
\caption{Simulation study: restricted-model non-invariance and imputation bias in non-selected patients (percentage points; mean over 1000 replicates per scenario, Monte Carlo standard error in brackets).}
\label{tab:sim}
\begin{tabular}{llrrrrr}
\toprule
Selection route & $\lambda_Z$ & Non-invariance & Residual non-invariance & \multicolumn{2}{c}{Imputation bias} & Difference \\
\cmidrule(lr){5-6}
 & & & & Distributional & Conditional mean & \\
\midrule
\multicolumn{7}{l}{\textit{Binary outcome}} \\
None & 0.0 & $-$0.01 (0.05) & $-$0.01 (0.05) & $-$0.01 (0.05) & $-$0.04 (0.05) & 0.00 (0.00) \\
Internal & 0.0 & 0.06 (0.05) & 0.06 (0.05) & 0.06 (0.05) & 0.03 (0.05) & $-$0.01 (0.00) \\
External & 0.0 & 5.22 (0.05) & 5.22 (0.05) & 5.22 (0.05) & 5.19 (0.05) & $-$0.00 (0.00) \\
Internal + external (moderate) & 0.0 & 3.44 (0.05) & 3.44 (0.05) & 3.44 (0.05) & 3.41 (0.05) & 0.00 (0.00) \\
Internal + external (strong) & 0.0 & 4.95 (0.05) & 4.95 (0.05) & 4.95 (0.05) & 4.93 (0.05) & $-$0.00 (0.00) \\
None & 0.8 & $-$0.07 (0.05) & $-$0.06 (0.05) & $-$0.08 (0.05) & $-$0.87 (0.05) & $-$0.00 (0.02) \\
Internal & 0.8 & 0.00 (0.05) & 0.01 (0.05) & $-$0.01 (0.05) & $-$0.79 (0.05) & $-$0.01 (0.02) \\
External & 0.8 & 5.25 (0.05) & 2.79 (0.05) & 5.28 (0.05) & 4.55 (0.05) & 0.03 (0.02) \\
Internal + external (moderate) & 0.8 & 3.34 (0.05) & 1.72 (0.05) & 3.34 (0.05) & 2.58 (0.05) & $-$0.01 (0.02) \\
Internal + external (strong) & 0.8 & 5.04 (0.05) & 2.62 (0.05) & 5.04 (0.05) & 4.32 (0.05) & $-$0.00 (0.02) \\
\midrule
\multicolumn{7}{l}{\textit{Time-to-event outcome (risk by 5 years, direct models)}} \\
Internal & 0.8 & 0.09 (0.07) & 0.05 (0.07) & 0.12 (0.08) & $-$1.10 (0.07) & 0.03 (0.03) \\
External & 0.8 & 8.59 (0.07) & 4.59 (0.07) & 8.57 (0.08) & 7.56 (0.07) & $-$0.01 (0.03) \\
Internal + external (moderate) & 0.8 & 5.87 (0.08) & 3.20 (0.08) & 5.84 (0.09) & 4.71 (0.08) & $-$0.03 (0.03) \\
Internal + external (strong) & 0.8 & 8.71 (0.09) & 4.68 (0.08) & 8.68 (0.09) & 7.56 (0.09) & $-$0.03 (0.03) \\
\bottomrule
\end{tabular}
\par\vspace{2pt}
{\footnotesize Residual non-invariance: non-invariance of the extended model evaluated with the true values of $Z$ in non-selected patients (oracle, available only by simulation). {\color{black}Difference: distributional imputation bias minus non-invariance, that is, the difference between the mean risks predicted in non-selected patients by the imputed extended model and by the restricted model; the theory predicts a value of zero.} About half of the patients were selected in each scenario; true values and further results are given in the Supplementary Material. 2 replicate(s) excluded because a direct model did not converge.}
\end{table}
\end{landscape}
}{}

\clearpage
\subsection*{Illustration using lactate in intensive care}

\subsubsection*{Data}

We illustrate the proposed framework using the PhysioNet/Computing in Cardiology Challenge 2012 dataset, a publicly available set of 12,000 adult intensive care unit (ICU) stays of at least 48 hours, extracted from the MIMIC-II database \autocite{silva2012predicting}. Each stay is described by demographic characteristics recorded at admission and by time-stamped measurements of vital signs and laboratory tests, and is linked to the outcome of in-hospital death. All predictors were taken from the first 24 hours after ICU admission. The data are de-identified and openly available; their secondary analysis did not require ethical approval.\\

In this setting, routine variables such as age, sex, type of ICU, heart rate, level of consciousness and basic laboratory tests are recorded for almost all patients, whereas blood lactate is measured selectively, when clinicians suspect hypoperfusion, circulatory failure or sepsis. Lactate is also a strong marker of mortality in critically ill patients, so that its selective measurement corresponds to the clinical setting described by DAG~(\ref{dag_0}).

\subsubsection*{Nested prediction models}

The outcome $Y$ was in-hospital death. The restricted model was based on predictors $X$ recorded in at least 95\% of stays: age, sex, type of ICU (coronary care, cardiac surgery recovery, medical, surgical), mean heart rate, lowest Glasgow Coma Scale score, and first values of blood urea nitrogen, creatinine, haematocrit and white blood cell count. The additional predictor $Z$ was the first lactate value (log scale), and $S$ indicated whether lactate was measured within the first 24 hours. Mechanical ventilation and arterial blood gas sampling within the first 24 hours were considered as observable manifestations of the underlying disease process ($W$), not included in the prediction models but potentially prompting lactate measurement. Stays with missing values of $X$ were excluded, leaving 11,362 stays, of which 5,752 (50.6\%) had lactate measured.\\

\subsubsection*{Statistical analysis}

The restricted model (logistic regression of $Y$ on $X$) and the extended model (logistic regression of $Y$ on $X$ and $Z$) were estimated in selected patients. The diagnostic tests described in the previous section were applied as follows. Selective measurement through the restricted-model predictors was assessed by a logistic regression of $S$ on $X$, and selective measurement through other manifestations of the disease process by adding $W$ to this model, both with likelihood ratio tests. Residual prognostic information of lactate (condition~\ref{a1}) was assessed in selected patients by a likelihood ratio test comparing the restricted and extended models. Restricted-model non-invariance was estimated by applying the restricted model estimated in selected patients to non-selected patients and comparing the mean predicted and observed risks of in-hospital death.\\

To illustrate the identity between imputation bias and restricted-model non-invariance, lactate was imputed in non-selected patients from its conditional distribution in selected patients, $\mathrm P(Z\mid X,S=1)$, estimated by a linear regression of log lactate on $X$. For each non-selected patient, the predicted risk was obtained by averaging the extended-model prediction over this conditional distribution, using 500 draws from the empirical residuals, in accordance with assumptions~\ref{a4} and~\ref{a5}. Imputation bias was defined as the mean predicted risk minus the observed risk in non-selected patients.\\

Confidence intervals and $p$-values for non-invariance and imputation bias were obtained from 2,000 bootstrap replicates, resampling selected and non-selected patients separately and re-estimating all models in each replicate.

\subsubsection*{Results}

Among the 11,362 ICU stays with complete restricted-model predictors, lactate was measured within the first 24 hours in 5,752 (50.6\%). Patients with lactate measured had a lower level of consciousness and were more often mechanically ventilated (78.1\% vs 44.1\%) or sampled for arterial blood gas (91.8\% vs 52.7\%) than patients without, and their in-hospital mortality was higher (18.4\% vs 10.3\%) (Table~\ref{tab:physionet1}).\\

\begin{table}[!ht]
\centering
\caption{Characteristics of ICU stays according to lactate measurement within the first 24 hours.}
\label{tab:physionet1}
\begin{tabular}{lrrr}
\toprule
 & Lactate not measured & Lactate measured & All \\
 & ($S=0$) & ($S=1$) & \\
\midrule
Number of stays & 5,610 & 5,752 & 11,362 \\
Age, years, mean & 65.5 & 63.7 & 64.6 \\
Male sex & 56.5\% & 56.2\% & 56.3\% \\
Medical ICU & 32.9\% & 37.0\% & 35.0\% \\
Cardiac surgery recovery unit & 23.9\% & 18.6\% & 21.2\% \\
Lowest Glasgow Coma Scale, median & 13 & 6 & 8 \\
Mechanical ventilation & 44.1\% & 78.1\% & 61.3\% \\
Arterial blood gas sampled & 52.7\% & 91.8\% & 72.5\% \\
In-hospital death & 10.3\% & 18.4\% & 14.4\% \\
\bottomrule
\end{tabular}
\end{table}

Lactate measurement was associated with the restricted-model predictors ($\chi^2=2{,}439.6$, 11 df, $p<0.001$) and, conditional on these, with mechanical ventilation and arterial blood gas sampling ($\chi^2=1{,}153.6$, 2 df, $p<0.001$), consistent with two routes of selection. In selected patients, lactate carried residual prognostic information beyond the restricted-model predictors (odds ratio 1.65 per unit of log lactate, $p<0.001$). Applied to non-selected patients, the restricted model estimated in selected patients predicted a mean risk of in-hospital death of 12.3\%, compared with an observed risk of 10.3\%, corresponding to a restricted-model non-invariance of 1.95 percentage points (95\% CI 0.75 to 3.17; $p=0.002$). The imputation bias was 2.00 percentage points (95\% CI 0.75 to 3.18), and its difference from restricted-model non-invariance was 0.05 percentage points (95\% CI $-0.23$ to 0.29), in agreement with the theoretical identity, the small difference reflecting the approximation of $\mathrm P(Z\mid X,S=1)$ by the imputation model (Table~\ref{tab:physionet2}).

\begin{table}[!ht]
\centering
\caption{Diagnostic tests and imputation bias in the PhysioNet/CinC Challenge 2012 data ($Z$: lactate; $S$: lactate measured within 24 hours).}
\label{tab:physionet2}
\begin{tabular}{p{8.6cm}p{3.9cm}r}
\toprule
Diagnostic & Estimate & $p$ \\
\midrule
Selective measurement through $X$ ($S$ on $X$) & $\chi^2=2{,}439.6$, 11 df & $<0.001$ \\
Selective measurement through $W$ given $X$ ($S$ on $X$ and $W$) & $\chi^2=1{,}153.6$, 2 df & $<0.001$ \\
Residual prognostic information of $Z$ (condition~\ref{a1}) & OR 1.65 per log unit & $<0.001$ \\
Restricted-model non-invariance, $\hat\Delta_{S=0}$ & 1.95 (0.75 to 3.17) & 0.002 \\
Imputation bias & 2.00 (0.75 to 3.18) & \\
Imputation bias $-$ non-invariance & 0.05 ($-0.23$ to 0.29) & \\
\bottomrule
\end{tabular}

\vspace{2pt}
{\footnotesize Non-invariance and imputation bias in percentage points of risk of in-hospital death in non-selected patients, with 95\% bootstrap confidence intervals.}
\end{table}

\subsection*{Implementation}

All analyses were conducted using R version 4.5.0 \autocite{R_2025}, with the packages survival and splines for the time-to-event analyses. The code for the simulation study and for the intensive care application is available as Supplementary Material. Claude Opus 5.5 (Anthropic; accessed September--November 2026) was used to edit text, write and edit R and LaTeX code, and draft parts of the Supplementary Material. All drafted text was revised by the author, and all code was reviewed, run, and checked by the author. The results of the intensive care application were reproduced with an independent script provided as Supplementary Material. The author takes full responsibility for the final content of the manuscript and Supplementary Material.

\section*{Discussion}

In this paper, we formalised selective measurement of a predictor and its potential consequences for clinical prediction models. Following Heckman's formulation of sample selection as a specification error \autocite{heckman1979sample}, we showed that non-invariance of the restricted model between selected patients and the target population decomposes into a component due to omission of the selectively measured predictor, resolved by its inclusion in the extended model, a residual component, and their interaction. This decomposition is structurally analogous to the Oaxaca--Blinder decomposition of differences in mean outcomes between groups \autocite{oaxaca1973,blinder1973}, here derived for differences in conditional outcome distributions between selected patients and the target population. Non-invariance requires selection through manifestations of the disease process beyond the restricted-model predictors; selection on these predictors alone, however strong, leaves both models invariant. Finally, we showed that imputation of the missing predictor from its conditional distribution in selected patients transfers restricted-model non-invariance to the imputed extended model as imputation bias. A simulation study illustrated this identity in finite samples, for binary and time-to-event outcomes, including when the extended model is itself non-invariant, and showed that the diagnostic tests kept their nominal level.\\

This framework corresponds to the structural definition of selection bias, in the setting of prediction models, with the selectively measured predictor playing the role of the exposure \autocite{Hernn2004}, under an external route of selection. It is characterised by two conditions: selective measurement of the predictor given the restricted-model predictors ($\neg$~\ref{a0}) and residual prognostic information beyond the restricted-model predictors (\ref{a1}). Strictly, the restricted model is not itself subject to selection, so its non-invariance does not formally constitute selection bias \autocite{lu2024selection}; the decomposition shows, however, that this observable non-invariance reflects the selection bias arising from the predictor-selection mechanism. These two conditions characterise restricted-model non-invariance under extended-model invariance. When the predictors partially measure the disease process, as in the simulation study, the extended model, which omits the other manifestations of the disease driving selection, is itself affected by selection, and non-invariance arises even when the selectively measured predictor is uninformative, entirely as residual non-invariance.\\

The results can also be framed as a question of identification. The extended-model prediction $\mathrm P(Y\mid X,Z)$ in the target population cannot in general be recovered from the observed data, since $Z$ is unobserved in non-selected patients. Restricted-model non-invariance, in contrast, is estimable from $Y$, $X$ and $S$ in all patients, and under the stated assumptions it equals the otherwise unobservable imputation bias. This links the present work to the treatment of sample selection as a specification problem \autocite{heckman1979sample} and to the identifiability of statistical relations across populations \autocite{pearl2011transportability}. In the terminology of the missing-data literature \autocite{vanBuuren2018}, when the extended model is invariant, imputation based on the selected population is unbiased under \ref{a0} and biased when $Z$ is missing not at random; the present framework shows that this bias is exactly identified by restricted-model non-invariance, and that it persists under \ref{a0} when the extended model is itself non-invariant. The diagnostic tests proposed here make these quantities assessable in routinely collected data, as illustrated in the intensive care application, where lactate measurement depended on both routes of selection and the imputed extended model reproduced the observed restricted-model non-invariance. When several external routes operate, as with mechanical ventilation and arterial blood gas sampling here, or diabetes and hypertension for the KFRE, decomposing restricted-model non-invariance across strata of these indicators would allow their respective contributions to be estimated; this is left for further work.\\

These results bear on the generalisability of prediction models to their intended target populations \autocite{steyerberg2009applications,sperrin2022targeted,ploddi2024scoping}. Generalisability and transportability both refer to the ability of a model to maintain its performance in new populations, transportability concerning target populations distinct from the development population \autocite{Justice1999}, a distinction not captured by that between internal and external validation \autocite{collins2024evaluation}. A model developed in a selected population may lack generalisability to its target population while remaining transportable to other populations selected through the same mechanism \autocite{pearl2011transportability,Major2019}. This is relevant in practice, as prediction models such as the Kidney Failure Risk Equation are frequently validated in populations defined by predictor availability rather than in the intended clinical target population \autocite{tangri2016multinational}.\\

From a practical perspective, imputation based on the conditional distribution of the predictor in selected patients makes the extended model applicable to non-selected patients \autocite{vanBuuren2018}, but it does not resolve the underlying non-identification: the extended model becomes computable at the cost of carrying restricted-model non-invariance forward as bias.  In the validation or application of an existing model, imputation of a selectively measured predictor is therefore not a remedy: conditional on the model's predictors, it reproduces restricted-model non-invariance exactly, while making it invisible. Conditioning the imputation on additional variables amounts to developing a new model, which requires its own validation and is exposed to the same problem whenever these variables are themselves incompletely or selectively recorded.\\

In terms of predictive performance, restricted-model non-invariance in non-selected patients is the calibration-in-the-large of the restricted model in that population, and the identity shows that an imputed extended model has the same calibration-in-the-large error. This does not preclude satisfactory predictive performance, but the error calls for recalibrating predictions separately in selected and non-selected patients, since a single recalibration of all patients cannot correct errors that differ between groups. Consequences for calibration across risk levels, discrimination and clinical utility remain to be studied.\\

In conclusion, nested prediction models under selective predictor availability provide a principled framework for understanding how selective measurement beyond the model's predictors induces non-invariance between selected patients and the target population, and how imputation transfers it into bias, with implications for the development, validation and use of clinical prediction models based on routinely collected health data.

\section*{Funding}
This work was supported by the National Institute for Health and Care Research (NIHR) (grant NIHR156488). The views expressed are those of the author and not necessarily those of the NIHR or the Department of Health and Social Care.

\section*{Conflict of interest}
None declared.

\section*{Data availability statement}
The PhysioNet/Computing in Cardiology Challenge 2012 data are publicly available from PhysioNet (\url{https://physionet.org/content/challenge-2012/1.0.0/}). The analysis and simulation code is available as Supplementary Material.

\appendix
\section*{Appendix}
\setcounter{section}{1}

\subsection{Residual prognostic information under restricted model non-invariance and extended model invariance}
\label{ap2}

In addition to the derivation of the conditional non-independence of $Z$ on $S$ given $X$ under restricted model non-invariance and extended model invariance given in the text, we provide here an alternative version of this derivation and formally derive non-independence of $Z$ on $Y$ given~$X$.\\

We assume that the extended model is invariant, that is,
\[
\mathrm{P}(Y\mid X,Z,S=1)
=
\mathrm{P}(Y\mid X,Z),
\]
while the restricted model is non-invariant,
\[
\mathrm{P}(Y\mid X,S=1)
\neq
\mathrm{P}(Y\mid X).
\]

Suppose that \(Z\) is independent of selection conditional on \(X\), i.e.
\[
Z \perp\!\!\!\perp S\mid X.
\]
Then
\[
\mathrm{P}(Z\mid X,S=1)=\mathrm{P}(Z\mid X),
\]
and therefore
\[
\begin{aligned}
\mathrm{P}(Y\mid X,S=1)
&=
\int
\mathrm{P}(Y\mid X,Z,S=1)\,
d\mathrm{P}(Z\mid X,S=1)
\\
&=
\int
\mathrm{P}(Y\mid X,Z)\,
d\mathrm{P}(Z\mid X,S=1)
\\
&=
\int
\mathrm{P}(Y\mid X,Z)\,
d\mathrm{P}(Z\mid X)
\\
&=
\mathrm{P}(Y\mid X).
\end{aligned}
\]

This contradicts the assumed non-invariance of the restricted model. Hence,

\[
Z \not\!\perp\!\!\!\perp S\mid X.
\]

Similarly, we can also derive non-independence of $Z$ on $Y$ given $X$.\\

Suppose that \(Z\) does not carry residual prognostic information, i.e.

\[
Z \perp\!\!\!\perp Y \mid X.
\]
Then
\[
\mathrm{P}(Y\mid X,Z)=\mathrm{P}(Y\mid X),
\]
and therefore
\[
\begin{aligned}
\mathrm{P}(Y\mid X,S=1)
&=
\int
\mathrm{P}(Y\mid X,Z,S=1)\,
d\mathrm{P}(Z\mid X,S=1)
\\
&=
\int
\mathrm{P}(Y\mid X,Z)\,
d\mathrm{P}(Z\mid X,S=1)
\\
&=
\int
\mathrm{P}(Y\mid X)\,
d\mathrm{P}(Z\mid X,S=1)
\\
&=
\mathrm{P}(Y\mid X).
\end{aligned}
\]

This contradicts the assumed non-invariance of the restricted model. Hence,
\[
Z \not\!\perp\!\!\!\perp Y\mid X.
\]

Therefore, under invariance of the extended model, non-invariance of the restricted model implies that \(Z\) both carries residual prognostic information and is selectively measured given $X$, i.e.:

\[
Z \not\!\perp\!\!\!\perp Y\mid X
\qquad\text{and}\qquad
Z \not\!\perp\!\!\!\perp S\mid X.
\]

\subsection{Non-invariance in the restricted model}
\label{ap3}

Non-invariance in the restricted model conditional on availability of \(Z\), can be expressed as

\[
\Delta_{X,S} =
\mathrm{P}(Y \mid X,S=1)-\mathrm{P}(Y \mid X)
\]

By integrating over the additional predictor \(Z\), we obtain
\[
\mathrm{P}(Y \mid X,S=1)
=
\int \mathrm{P}(Y \mid X,Z,S=1)\,d\mathrm{P}(Z \mid X,S=1),
\]
and
\[
\mathrm{P}(Y \mid X)
=
\int \mathrm{P}(Y \mid X,Z)\,d\mathrm{P}(Z \mid X).
\]

Therefore,
\[
\Delta_{X,S}
=
\int \mathrm{P}(Y \mid X,Z,S=1)\,d\mathrm{P}(Z \mid X,S=1)
-
\int \mathrm{P}(Y \mid X,Z)\,d\mathrm{P}(Z \mid X).
\]

After adding and subtracting both cross-terms:

\[
\int \mathrm{P}(Y \mid X,Z, S=1)\,d\mathrm{P}(Z \mid X)
\]

and

\[
\int \mathrm{P}(Y \mid X,Z)\,d\mathrm{P}(Z \mid X,S=1)
\]

we obtain

\[
\begin{aligned}
\Delta_{X,S}
=&
\int
\mathrm{P}(Y \mid X,Z)\,
\left[
d\mathrm{P}(Z \mid X,S=1)
-
d\mathrm{P}(Z \mid X)
\right]
\\
&+
\int
\left[
\mathrm{P}(Y \mid X,Z,S=1)
-
\mathrm{P}(Y \mid X,Z)
\right]
d\mathrm{P}(Z \mid X)
\\
&+\int
\left[
\mathrm{P}(Y \mid X,Z,S=1)
-
\mathrm{P}(Y \mid X,Z)
\right]\,\left[
d\mathrm{P}(Z \mid X,S=1)
-
d\mathrm{P}(Z \mid X)
\right].
\end{aligned}
\]

For improved interpretability, terms of this expression can be written as

\[
\begin{cases}
\Delta^{\rm exp}_{Z \mid X}
=
\int
\mathrm{P}(Y \mid X,Z)\,
d \delta_{Z \mid X}
\\
\Delta^{\rm res}_{Z , X}
=
\int
\delta_{X,Z}\,
d\mathrm{P}(Z \mid X)
\\
\Delta^{\rm int}_{Z , X}
=
\int
\delta_{X,Z}
\,
d\delta_{Z \mid X},
\end{cases}
\]

and finally, we obtain

\[
\Delta_{X,S}
=
\Delta_{Z\mid X}^{\mathrm{exp}}
+
\Delta_{Z , X}^{\mathrm{res}}
+
\Delta_{Z , X}^{\mathrm{int}}.
\]

\subsection{Implications for prediction model validation}
\label{ap4}

We express the difference between the predictive distribution of the outcome after imputation and the true unknown predictive distribution of the outcome in the non-selected population as

$$
\delta^{\mathrm{imp}}_{Z\mid X,S=0}
=
\hat P(Y\mid X,S=0)
-
P(Y\mid X,S=0).
$$

The first term can be integrated over \(\hat Z\):

$$
\hat P(Y\mid X,S=0)
=
\int
P(Y\mid X,\hat Z, S=0)
\,dP(\hat Z\mid X,S=0).
$$

Using \ref{a4} and \ref{a5} we obtain

$$
\hat P(Y\mid X,S=0)
=
\int
P(Y\mid X,Z,S=1)
\,dP(Z\mid X,S=1) = P(Y\mid X,S=1)
$$

and therefore

$$
\delta^{\mathrm{imp}}_{Z\mid X,S=0}
=
P(Y\mid X,S=1)
-
P(Y\mid X,S=0).
$$

To express the imputation bias over the target population, we use

$$
P(Y\mid X)
=
\pi_X P(Y\mid X,S=1)
+
(1-\pi_X)P(Y\mid X,S=0),
$$

where

$$
\pi_X=P(S=1\mid X).
$$

Hence,

$$
P(Y\mid X,S=1)
-
P(Y\mid X)
=
(1-\pi_X)
\left[
P(Y\mid X,S=1)
-
P(Y\mid X,S=0)
\right].
$$

The imputation bias over the target population is

$$
\Delta^{\mathrm{imp}}_{Z\mid X}
=
\pi_X
\delta^{\mathrm{imp}}_{Z\mid X,S=1}
+
(1-\pi_X)
\delta^{\mathrm{imp}}_{Z\mid X,S=0}.
$$

However, since imputation is not performed in selected patients,

$$
\delta^{\mathrm{imp}}_{Z\mid X,S=1}=0.
$$

so that

$$
\Delta^{\mathrm{imp}}_{Z\mid X}
=
(1-\pi_X)
\delta^{\mathrm{imp}}_{Z\mid X,S=0}.
$$

Substituting the previous result yields

\[
\Delta^{\mathrm{imp}}_{Z\mid X}
=
(1-\pi_X)
\left[
P(Y\mid X,S=1)
-
P(Y\mid X,S=0)
\right]
\]

\[
=
P(Y\mid X,S=1)
-
\pi_X P(Y\mid X,S=1)
-
(1-\pi_X)P(Y\mid X,S=0)
\]

\[
=
P(Y\mid X,S=1)
-
\left[
\pi_X P(Y\mid X,S=1)
+
(1-\pi_X)P(Y\mid X,S=0)
\right].
\]

Since

\[
P(Y\mid X)
=
\pi_X P(Y\mid X,S=1)
+
(1-\pi_X)P(Y\mid X,S=0),
\]

we finally obtain

\[
\Delta^{\mathrm{imp}}_{Z\mid X}
=
P(Y\mid X,S=1)
-
P(Y\mid X).
\]

Therefore,

$$
\Delta^{\mathrm{imp}}_{Z\mid X}
=
\Delta_{X,S}.
$$

Using the decomposition of restricted-model non-invariance derived in Appendix A.2,

$$
\Delta_{X,S}
=
\Delta^{\mathrm{exp}}_{Z\mid X}
+
\Delta^{\mathrm{res}}_{Z,X}
+
\Delta^{\mathrm{int}}_{Z,X},
$$

we finally obtain

$$
\Delta^{\mathrm{imp}}_{Z\mid X}
=
\Delta^{\mathrm{exp}}_{Z\mid X}
+
\Delta^{\mathrm{res}}_{Z,X}
+
\Delta^{\mathrm{int}}_{Z,X}
.
$$

Under the assumption that \(\delta_{X,Z}\) does not depend on \(Z\), the interaction term vanishes, and

$$
\Delta^{\mathrm{imp}}_{Z\mid X}
=
\Delta^{\mathrm{exp}}_{Z\mid X}
+
\Delta^{\mathrm{res}}_{Z,X}
=
\Delta_{X,S}.
$$

\subsection{The imputation bias at the conditional level $\delta_{X,Z,S=0}^{ \rm imp}$ in the target population}
\label{ap6}

At the conditional level,

\[
\delta_{X,Z,S=0}^{\rm imp}
=
\textrm{P}(Y\mid X,\hat Z,S=0)
-
\textrm{P}(Y\mid X,Z,S=0).
\]

Under (\ref{a5}),

\[
\delta_{X,Z,S=0}^{\rm imp}
=
\textrm{P}(Y\mid X,Z,S=1)
-
\textrm{P}(Y\mid X,Z,S=0).
\]

Since

\[
\textrm{P}(Y\mid X,Z)
=
\pi_{X,Z}\,\textrm{P}(Y\mid X,Z,S=1)
+
(1-\pi_{X,Z})\,\textrm{P}(Y\mid X,Z,S=0),
\]

we obtain for  \(P(S=0\mid X,Z)>0\),

\[
\delta_{X,Z,S=0}^{\rm imp}
=
\frac{
\textrm{P}(Y\mid X,Z,S=1)
-
\textrm{P}(Y\mid X,Z)
}
{1-\pi_{X,Z}}
=
\frac{\delta_{X,Z}}{(1-\pi_{X,Z})}.
\]

Considering

\begin{align*}
\delta_{X,Z}^{\rm imp} &= \pi_{X,Z} \delta_{X,Z,S=1}^{\rm imp} + (1-\pi_{X,Z}) \delta_{X,Z,S=0}^{\rm imp},
\end{align*}

and since $\hat{Z} = Z$ in selected patients, we have

\begin{align*}
\delta_{X,Z,S=1}^{\rm imp}
&=
\textrm{P}(Y\mid X,\hat Z,S=1)
-
\textrm{P}(Y\mid X,Z,S=1)\\
&=
\textrm{P}(Y\mid X,Z,S=1)
-
\textrm{P}(Y\mid X,Z,S=1)\\
&= 0
\end{align*}

and finally

\begin{align*}
\delta_{X,Z}^{\rm imp}
&= (1-\pi_{X,Z}) \delta_{X,Z,S=0}^{\rm imp}\\
&= (1-\pi_{X,Z}) \frac{\delta_{X,Z}}{(1-\pi_{X,Z})}\\
&= \delta_{X,Z}.
\end{align*}

\printbibliography

\end{document}